\documentclass[10pt,journal,compsoc]{IEEEtran}

\ifCLASSOPTIONcompsoc
\usepackage[nocompress]{cite}
\else
\usepackage{cite}
\fi

\ifCLASSINFOpdf

\else

\fi

\usepackage{graphicx}
\usepackage{xcolor}
\usepackage{float}

\begin{document}
	
	\title{Crypto Makes AI Evolve}

\author{\IEEEauthorblockN{Behrouz Zolfaghari\IEEEauthorrefmark{2},  Elnaz Rabieinejad\IEEEauthorrefmark{2}, Abbas Yazdinejad\IEEEauthorrefmark{2}, Reza~M. Parizi\IEEEauthorrefmark{3}, Ali Dehghantanha\IEEEauthorrefmark{2}
		}
	\IEEEauthorblockA{
	 \\
		\IEEEauthorrefmark{2}Cyber Science Lab, School of Computer Science, University of Guelph,
		Ontario, Canada  \\
	behrouz@cybersciencelab.org, erabiein@uoguelph.ca ,ayazdine@uoguelph.ca, adehghan@uoguelph.ca\\
					\IEEEauthorrefmark{3}College of Computing and Software Engineering, Kennesaw State University, GA, USA\\  rparizi1@kennesaw.edu  \\
							}}


	
	\IEEEtitleabstractindextext{%
		\begin{abstract}
			Adopting cryptography has given rise to a significant evolution in Artificial Intelligence (AI). This paper studies the path and stages of this evolution. We start with reviewing existing relevant surveys, noting their shortcomings, especially the lack of a close look at the evolution process and solid future roadmap. These shortcomings justify the work of this paper. Next, we identify, define and discuss five consequent stages in the evolution path, including Crypto-Sensitive AI, Crypto-Adapted AI, Crypto-Friendly AI , Crypto-Enabled AI, Crypto-Protected AI. Then, we establish a future roadmap for further research in this area, focusing on the role of quantum-inspired and bio-inspired AI.
		\end{abstract}
		
		}

	\maketitle

	\IEEEdisplaynontitleabstractindextext
	
	\IEEEpeerreviewmaketitle

	\section{Introduction}\label{sec:introduction}
	
	
	 In recent decades, Artificial intelligence (AI) has been of great interest to the research community \cite{Introooo-Jour001}\cite{Introooo-Jour002,a1}. AI refers to the use of advanced mathematics and layers of complex algorithms in tasks that would typically require human intelligence, such as \textcolor{black}{classification}. AI is used in a wide variety of \textcolor{black}{environments} ranging from edge and cloud computing \cite{Introooo-Jour007} to aerospace engineering \cite{Introooo-Jour008}, remote sensing \cite{Introooo-Jour009}, meteorology \cite{Introooo-Jour011}, and medical systems \cite{Introooo-Jour010}.
	 
	 Different design objectives such as energy-efficiency \cite{Introooo-Jour003,a2}, stability \cite{Introooo-Jour004}, computational performance \cite{Introooo-Jour005,a3} and resource constraint awareness \cite{Introooo-Jour006}  are considered in the design of AI-based systems. \textcolor{black}{However}, security is probably the most critical design objective in such systems \cite{Introooo-Jour012}\cite{Introooo-Jour014,abc}. \textcolor{black}{Among all security measures, which frequently appear in the ecosystem of AI, this paper focuses on cryptography.}
	 
\begin{figure}[h]
	    \centering
	    \includegraphics[width=0.85\linewidth,keepaspectratio]{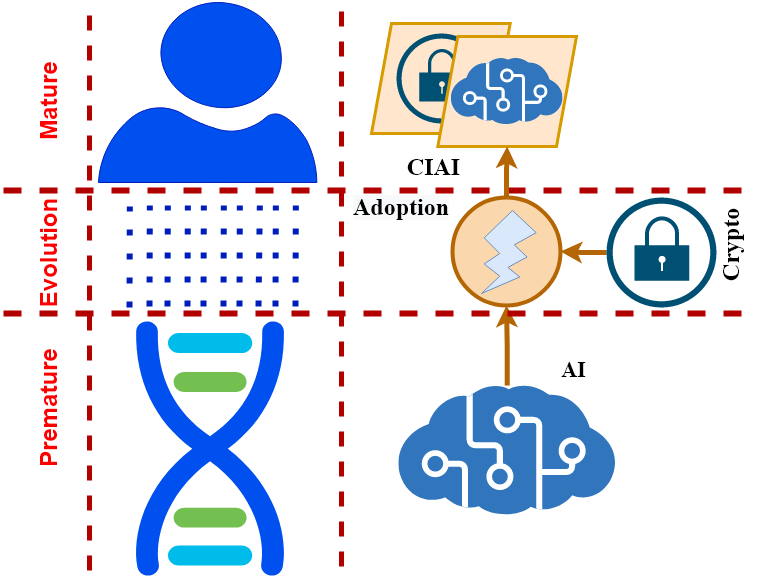}
	    \caption{\textcolor{black}{Crypto Makes AI Evolve}}
	    \label{fig:AI_on_Crypto}
	\end{figure}

	 
	 \textcolor{black}{Cryptography is the art and science of leveraging algorithms, mathematical problems and structures, secret keys, and complex transforms to keep data confidential while being stored or transmitted.} Cryptography is used in a broad variety of applications via serving to different \textcolor{black}{security controls} such as information hiding \cite{Crypt-Jour007-1}, authentication \cite{Crypt-Jour008-1}, privacy \cite{Crypt-Jour009-1}. Modern cryptography is connected with several branches of science and technology including, but not limited to quantum computing \cite{Acc-Jour-New001}, information theory \cite{New-Entropy-Jour001}\cite{Entropy-Jour001}, chaos theory \cite{ASI-Chaot-Enc} and  hardware technology \cite{Acc-Book2,a4} in its ecosystem. \textcolor{black}{The adoption of Cryptography has played a critical role in the evolution of AI \cite{ASI-Ware-Peace} in a way that it cannot be separated from AI anymore. Figure  \ref{fig:AI_on_Crypto} illustrates this role.}
	  
	 \textcolor{black}{In Figure \ref{fig:AI_on_Crypto}, the overlapping parallelograms represent evolved AI after adopting crypto, to which \textit{\textbf{Crypto-Influenced AI (CIAI)}} in the rest of this paper. We use the general term CIAI to refer to those AI methods, models, algorithms and concepts that have been affected by cryptography.}

	 \textcolor{black}{There are numerous research works focusing on the application of cryptography in secure Machine Learning (ML) and Deep Learning (DL). However, the role of cryptography in the evolution of AI has not been studied in-depth. In fact, the current literature does not provide a thorough understanding and a comprehensive perspective on the impact of cryptography on the present and the future of AI. This review is an attempt to address this gap, which makes it pertinent to conduct a survey CIAI. Such a survey can highlight existing trends and shed light on less-studied aspects. Furthermore, it simplifies future research and investigation.}
	 
\subsection{\textcolor{black}{Objectives}} \label{Obj} 
	 
	 In this review, we attempt to answer the following questions via exploring the evolution path and stages of CIAI \textcolor{black}{as well as} its future roadmap.
 \begin{itemize}
	\item What stages should AI go through so that it adopts the Crypto in itself?
	\item What additional capabilities AI will obtain in these stages?
	\item \textcolor{black}{What enabling technologies will play roles in the future of CIAI?}
	\item \textcolor{black}{What is the future expected to hold for CIAI given existing trends in AI and cryptography?}
	\end{itemize}




\subsection{\textcolor{black}{Achievements}}\label{Ach}

\textcolor{black}{The achievements of this survey can be listed as follows.}

\begin{enumerate}
    \item We recognize and define the following five different evolution stages of CIAI.
	  
	  \begin{itemize}
	      
	      \item \textbf{Crypto-Sensitive AI(CSAI)}:  Capable of detecting (benign/malign) encryption in ciphertext without decryption.
	      
	      \item \textbf{Crypto-Adapted AI (CAAI)}: Capable of processing and analyzing encrypted input data and returning the result in encrypted form.
	    
	      \item \textbf{Crypto-Friendly AI (CFAI)}: Considering that issues such as security and privacy threaten the use of data to train AI-based systems, CFAI is capable of being trained on encrypted datasets.
	      
	      \item \textbf{Crypto-Enabled AI (CEAI)}: Capable of using AI inheritance features, such as making complex ciphertext, to improve encryption and decryption in cryptographic functions.
	      
	      \item \textbf{Crypto-Protected AI (CPAI)}: AI and crypto get Interwoven. AI models get secured by cryptographic mechanisms, and cryptography gets more robust using AI features.
	      
	  \end{itemize}
	  
    \item We review current AI trends and then predict the influence of these trends on the future of CIAI. Moreover, we develop a future roadmap for further research in this area. Hereon, we anticipate the impact of bio-inspired and quantum-inspired AI on the future of research in this area.
\end{enumerate}

 \subsection{\textcolor{black}{Organization}}	 
	The rest of this paper is organized as follows. Section \ref{ExSurv} reviews existing relevant surveys in order to highlight their shortcomings and \textcolor{black}{clarify} our motivations for the work of this paper. Section \ref{State} discusses the evolution path and stages. Section \ref{Fut} develops a future roadmap for further research in this area, and \textcolor{black}{lastly}, Section \ref{Conc} concludes the paper.

\section{Existing Surveys}\label{ExSurv}

\textcolor{black}{This section studies existing relevant surveys to shed light on their shortcomings, which motivate the survey reported in this paper.}	

\subsection{Surveys on Secure AI}
	
Secure AI refers to methods that can be used to reduce problems in the application of AI methods, such as methods for privacy-preserving machine learning and inference. Secure AI has been of interest to many researchers in recent years. As a result, there have been several surveys published on this topic. Most such surveys have focused on a specific security aspect such as privacy or resilience in a specific AI tool such as neural networks, Machine Learning (ML), or Deep Learning (DL) \cite{a6,a7}. For example, the authors of \cite{New-New-Jour001} have studied collaborative (distributed) DL from a privacy point of view. They have discussed general cryptographic algorithms and other techniques for preserving privacy in collaborative DL. Resilient ML has been studied in another survey with a focus on applications in networked cyber-physical systems \cite{New-New-Jour002}. The latter survey argues that resilience is a necessary property for ML algorithms used in secure cyber-physical systems. Finally, robust ML for secure healthcare systems has been reviewed in \cite{New-New-Jour003}. Various privacy-preserving scenarios in healthcare have been investigated in \cite{New-New-Jour003} along with potential robust ML methods for each scenario.

Moreover, a survey of privacy-preserving DL methods based on multiparty computation and data encryption has been presented in \cite{New-New-4} with a focus on privacy in the training and inference phases. In \cite{New-New-4}, robust DL methods have been categorized depending on whether they use linear or nonlinear analyses as building blocks.

Security and privacy issues in DL have been summarized in \cite{New-New-Jour007} along with associated attack and defense methods, focusing on adversarial attacks. In addition, the mentioned survey discusses some related challenges and poses some open problems regarding privacy in DL.

Another review focusing on secure AI has been reported in \cite{New-New-Jour005}, which studies attacks on DL models. This survey particularly focuses on four types of attacks, namely  \textit{model inversion attack}, \textcolor{black}{\textit{model extraction attack, }}\textit{adversarial attack} ,and \textit{poisoning attack} along with related essential workflow as well as attack goals and adversary capabilities.
Security threats against ML, along with the related defensive techniques, have been reviewed in \cite{New-New-Jour006}. The authors of \cite{New-New-Jour006} classify existing countermeasures against ML attacks into \textit{security assessment mechanisms}, \textit{training phase countermeasures}, \textit{testing phase countermeasures}, \textit{inferring phase countermeasures}, \textit{data security mechanisms}, and \textit{privacy preservation}. Moreover, they highlight some related topics on which there is room for future research. A brief overview of emerging research problems related to secure ML, along with a survey on current progress in this area, has been presented in another research work \cite{New-New-Conf001}.

A broad range of attacks \textcolor{black}{ are }commonly conducted against ML has been discussed by He, et al. along with the related corresponding countermeasures \cite{New-New-Jour005}. In another overview on secure ML, Liao, et al. highlighted adversarial attacks among the most critical threats against ML models \cite{New-New-Conf001}.  

Zhang, et al. argued that privacy is a major concern when it comes to DL \cite{New-New-4} due to significant dependency on large datasets. They discussed how progress is made in the design of privacy-preserving DL and how further research can be conducted in this area. In another survey, Antwi-Boasiako, et al. observed that the idea of Collaborative DL (CDL) is more about the training stage than the inference stage \cite{New-New-Jour001}. They showed in their review that privacy preservation as a main component of CDL is improved as collusion resistant and quantum robustness are investigated by researchers. Moreover, they concluded that homomorphic encryption holds an excellent promise for privacy-preserving DL, provided that the computation and communication costs are efficiently managed. Olowononi, et al. highlighted resilience as a significant need in secure cyber-physical systems considering the increasing vulnerabilities caused by the gap between sensors and actuators \cite{New-New-Jour002}. They remarked that although AI and ML can serve the security of cyber-physical systems, they need resilience themselves.

 Qayyum, et al. investigated the vulnerabilities of ML and DL models to adversarial attacks and recommended some defensive approaches, including but not limited to examining the raw data and understanding the model and dataset's limitations, as well as continuous monitoring and testing  \textcolor{black}{\cite{New-New-Jour003}}. In addition, there is another review on the vulnerability of ML models to adversarial attacks \cite{New-New-Jour006}. The authors of the latter report highlighted adversarial attacks as the most emerging threats against ML and the assessment of the vulnerability of ML to these attacks as a prominent trend in research on this area.

\subsection{Surveys on the Role of Cryptography}

AI techniques can be applied to improve cryptography in various ways. In this way, ML and DL have been adopted for encryption and cryptoanalysis. Cryptography is the practice of converting plaintext to a ciphertext to keep it confidential. The ciphertext should ideally avoid creating patterns to protect data and prevent hacking. Furthermore, AI's ability to generate complex patterns is a good option for cryptography and particularly for cipher generation. For example, Arnaut, et al. showed that the multi-objective genetic training algorithm performed was best able to classify SSH and non-SSH traffic when compared to C4.5 and k-means algorithms on the same network as the training data \cite{Crypt-In-AI-Conf062}. However, they observed that the C4.5 training algorithm performed better than the other algorithms when tested on a different network from the training data.

In another work, Wang, et al. acknowledge DL becoming desirable for classifying encrypted network traffic and discuss current work in the area \cite{Crypt-In-AI-Jour061}. Indeed, they review most of the existing work according to dataset selection, model input design, and model architecture until they propose some noteworthy DL-based mobile services traffic classification challenges. Also, in \cite{Surv-Surve-Conf001}, 30 state-of-the-art image encryption classification models have been reviewed. The researchers found that models focus either on the security of the techniques or on the optimization of the model. Their optical security (PNO) system aims to resolve this issue. They conclude that using neural network-based encryption techniques can increase security.

 A review of the literature on symbiotic relationships between machine learning and cryptography applications is presented in \cite{Surv-Surve-Conf002}. The authors demonstrate that cryptographic algorithms can successfully be applied to ML issues, while DL algorithms are appropriate for cryptography. Lastly, they pay particular attention to the area of adversarial robustness. 
Kiya also discusses progress and challenges in privacy-preserving encryption and machine learning  \cite{New-New-Conf002}. This paper discusses compression-then-compression (EtC) systems and learnable encryption systems, which have been proposed for encrypted image encryption. 

\textcolor{black}{Table \ref{table:surveys} summarizes the surveys discussed in this section in order to facilitate identifying their shortcomings and comparing them to our work in this paper.}

\begin{table}[]
\centering
	\caption{\textcolor{black}{A Summary of Existing Surveys}}
\begin{tabular}{|c|c|c|c|c|c|}
\hline
\textbf{Survey} & \textbf{Year} &\textbf{Crypto} &\textbf{General AI} & \textbf{Roadmap} \\ \hline\hline
\cite{Crypt-In-AI-Conf062}    & 2011          & Yes                         & No               & No                          \\ \hline
\cite{New-New-Conf001}     & 2011          & No                          & No              & No          \\ \hline
\cite{New-New-Jour005}     & 2019          & No                          & No             & Yes          \\ \hline
\cite{New-New-Jour006}     & 2018          & No                          & No              & Yes          \\ \hline
\cite{New-New-Jour001}     & 2021          & No                          & No              & No         \\ \hline
\cite{Crypt-In-AI-Jour061}     & 2019          & yes                          & No              & No          \\ \hline
\cite{New-New-Jour002}     & 2021          & No                          & No              & No           \\ \hline
\cite{New-New-Jour003}     & 2021          & No                          & No              & Yes           \\ \hline
\cite{New-New-4}    & 2020          & No                         & No               & Yes          \\ \hline
\cite{New-New-Jour007}    & 2020          & No                          & No              & Yes          \\ \hline
\cite{Surv-Surve-Conf001}     & 2020          & Yes                         & No               & No          \\ \hline
\cite{Surv-Surve-Conf002}    & 2020          & Yes                         & No               & No                \\ \hline
\cite{New-New-Conf002}     & 2020          & Yes                         & No              & No          \\ \hline
\end{tabular}
\label{table:surveys}
\end{table}

 \textcolor{black}{The first column in each row of Table \ref{table:surveys} cites one of the surveys discussed earlier. The second column contains the publication year of the cited survey. This column helps identify outdated surveys. The third column indicates whether the survey focuses specifically on cryptography or other security aspects and mechanisms. The fourth column contains a "Yes" if the survey studies the general concept of AI (instead of focusing on ML, DL, neural networks, etc.). The fifth column shows the existence or the lack of a future roadmap in the survey.}
 
\subsection{\textcolor{black}{Motivations}}

\textcolor{black}{The literature comes with some surveys on secure AI or the role of AI in security, a few of which were studied earlier in this section. However, as seen in Table \ref{table:surveys},} some of them are too outdated for such a fast-moving research area \cite{Crypt-In-AI-Conf062}\cite{New-New-Conf001}. Others do not focus on the role of cryptography \cite{New-New-Jour005}\cite{New-New-Jour006}. \textcolor{black}{Further, some existing surveys focus on specific aspects of AI instead of covering the whole branch of science. }Some existing surveys fail to give directions for future research \cite{New-New-Jour001}\cite{Crypt-In-AI-Jour061}. Most importantly, to the best of our knowledge, there is no survey focusing on the evolution of AI under the impact of cryptography. \textcolor{black}{This makes it pertinent to conduct a comprehensive surveys with the objectives discussed in Subsection \ref{Obj} and the achievements mentioned in subsection \ref{Ach}.}



\section{The Evolution Path and Stages}\label{State}
In today's modern world, where innovative technologies are part of people's daily lives and are connected to their personal information, security and privacy are one of the most critical concerns. Hackers try to discover system vulnerabilities and exploit them in various ways every day. Numerous reports of malicious system activity, even with cryptography, indicate that traditional cryptographic methods are not immune to new attacks. So over time, researchers have been looking for ways to upgrade cryptographic algorithms and reduce systems security vulnerabilities.

AI is an extensive scientific discipline that can improve cryptography and cryptoanalysis by adopting complex learning, reasoning, and self-correction processes. AI can enhance cryptography in several ways. For example, because of its ability to generate complex patterns, AI can produce more complex ciphertexts by maximizing diffusion and confusion. In another example, AI tries to improve the cryptographic algorithm by examining the power of algorithms and the points that hackers use to break the encryption. Therefore, according to the different roles AI can play in cryptography, the five stages of the evolution of CIAI can be defined as CSAI, CAAI, CFAI, CEAI, and CPAI.  AI should go through these stages to adopt the Crypto in itself. 

\begin{figure}[h!]
	    \centering
	    \includegraphics[width=0.9\linewidth,keepaspectratio]{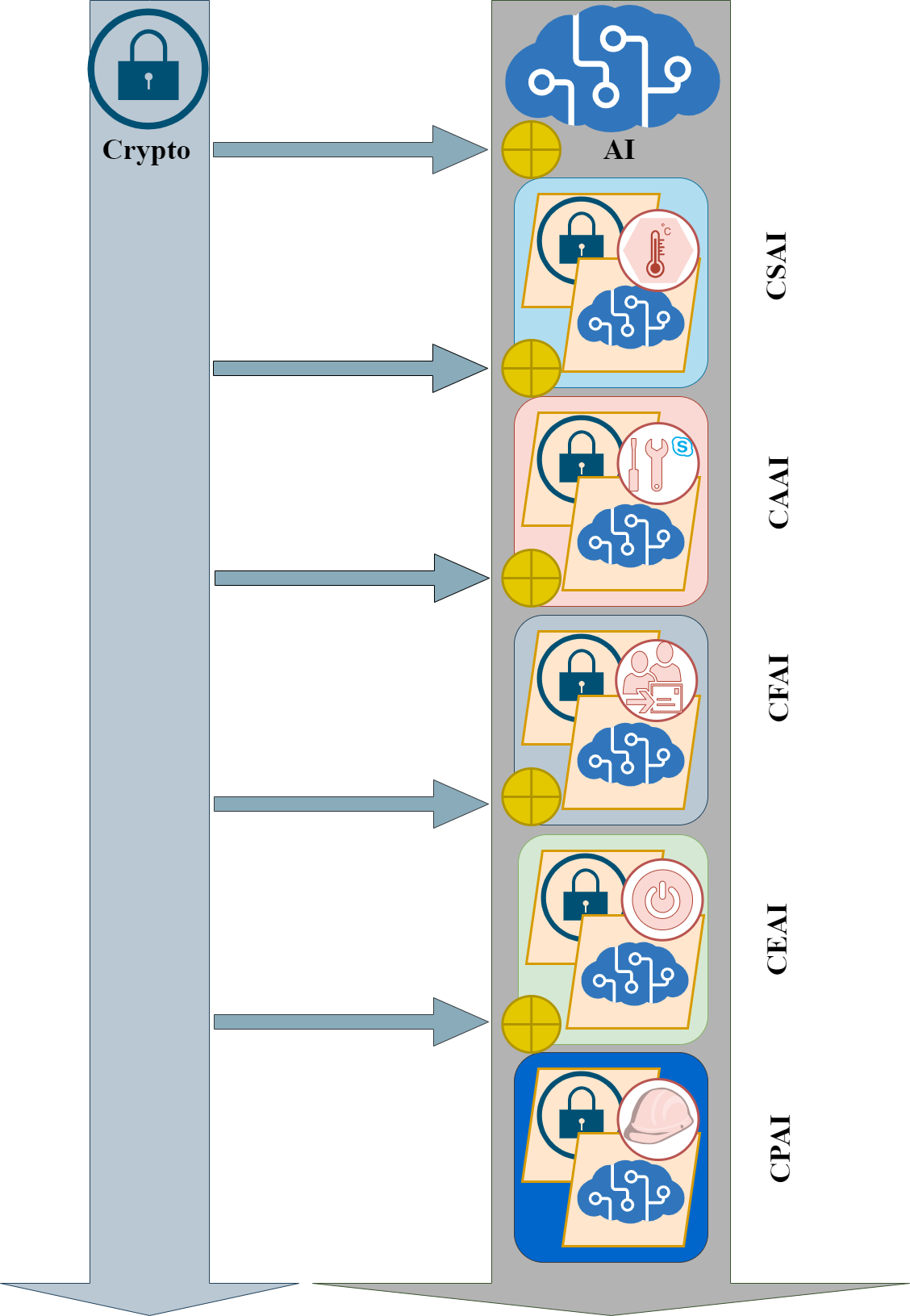}
	    \caption{The Evolution Stages of CIAI}
	    \label{fig:evolution_nn_cryptography}
	\end{figure}

\textcolor{black}{Figure} \ref{fig:evolution_nn_cryptography} is a visual representation of the stages of the evolution of CIAI. Bringing Crypto and AI together to meet these steps is important. We can use Crypto-AI applications to detect malign encryption, process encrypted data, train AI models using encrypted data, advance encryption, and protect AI systems. The evaluation should be based on each path and its features and characteristics due to the application-specific. In the following, we will discuss each of these stages. 
	
\subsection{CSAI}
	
As security and privacy concerns grow, the usage of encryption in data has increased significantly. While encryption improves data security and privacy, it also covers hackers' malicious and illegal behaviors on the data. Detecting malicious behaviour and attacks on encrypted data is difficult due to the complexity of ciphertext. Detection of attacks on encrypted data is divided into two general categories: \textit{post-decryption detection} and \textit{non-decryption detection}. Since the post-decryption detection approach is time and energy-consuming, its application is not suggested. As a result, researchers have been looking for an effective way to detect attacks without the need for decryption \cite{Crypt-In-AI-Jour064,New-New-Conf008, New-New-Conf0011}.As a non-decryption detecting method, AI shows remarkable performance in detecting benign encrypted data \cite{yang2021malicious}.

\begin{figure}[h]
	    \centering
	    \includegraphics[width=0.9\linewidth,keepaspectratio]{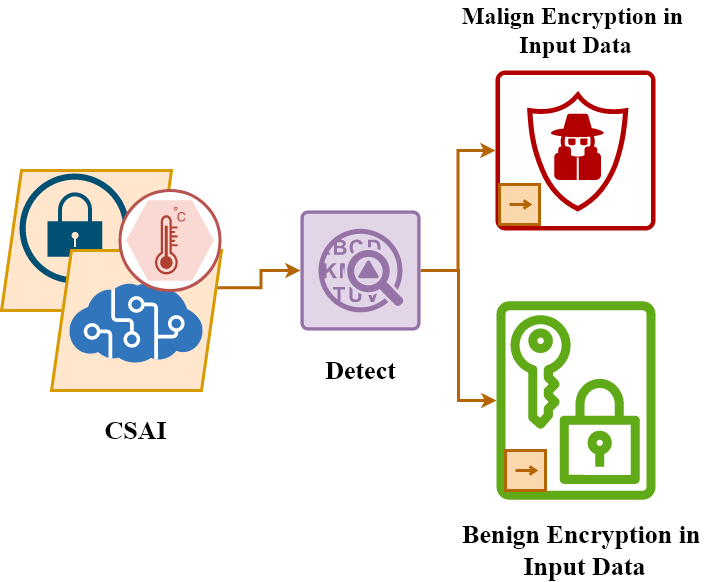}
	    \caption{\textcolor{black}{CSAI Capable of Detecting Benign and Malign Encryption in Input Data}}
	    \label{fig:Crypt-Sens-AI}
	\end{figure}

 \textcolor{black}{As shown in Figure} \ref{fig:Crypt-Sens-AI}, CSAI refers to the type of AI that is not only used to detect maliciously encrypted data but also, in some cases, is used for detecting benign encrypted data. For example,  Pastor, et al. showed that ML could be used to detect encrypted cryptomining malware in a real-time environment \cite{Crypt-In-AI-Jour064}. They also identified network flow features that can be used in a real-time environment for detection. In response to adversarial attacks via encrypted network packets, Bazuhair, et al. developed a method to generalize further a convolutional neural network \cite{New-New-Conf008}. They then extracted known relevant features and encoded them to train the DL model. Miller, et al. described a framework that can identify whether a packet is from a VPN or an original IP in real-time via a multilayer perceptron neural network \cite{New-New-Conf009}. Yu, et al. developed a method for detection of malicious encrypted traffic via a neural network for detection and used multilayer autoencoder networks to extract features \cite{New-New-Conf0010}. They also investigated the influence of dataset imbalance on outcomes in encrypted traffic detection. Moreover, Wright, et al. created a system that classifies components of a program as either related to or not related to cryptography \cite{New-New-Conf0011}. Using an artificial neural network, the authors classified functional blocks from a disassembled program based on cryptography-relation
 
In another research, the authors used graph embeddings and ML to detect cryptography misuse in source code \cite{New-New-Conf0034}. The authors compare node2vec, graph embedding techniques, and Bag of Graphs as embedding generators of source code graph representations. The main goal is to detect cryptography misuses in source code by combining these techniques with ML models. Finally, they showed that Bag of Graphs outperformed node2vec in this task; also, both techniques outperformed previously evaluated tools.

To summarize, different AI methods, including ML and DL, are capable of seeing attacks through encryption. In other words, the knowledge that AI methods gain from extracting and analyzing relevant features causes malign encrypted data to be identified. Furthermore, CSAI makes it possible to efficiently detect the malicious behaviour of hackers hiding behind complex encryption.

\subsection{CAAI}
\begin{figure}[h]
	    \centering
	    \includegraphics[width=0.9\linewidth,keepaspectratio]{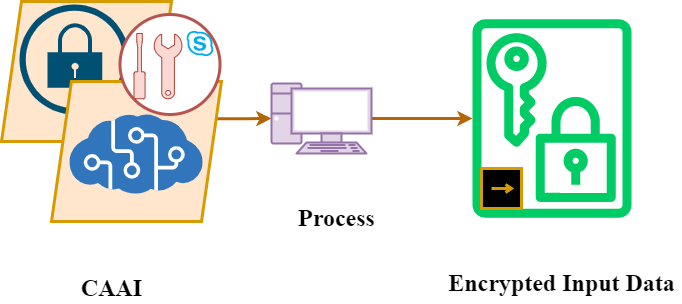}
	    \caption{CAAI Capable of Processing Encrypted Input Data}
	    \label{fig:Crypt-Adap-AI}
	\end{figure}

Analyzing and processing privacy-sensitive data has always been a concern for customers. Traditional data analysis methods only can process the plain text form of data and information. Converting the encrypted data to plain text not only threatens data privacy but also is cost-consuming. As we can see in \textcolor{black}{Figure} \ref{fig:Crypt-Adap-AI}, different AI methods, including ML and DL, are effective and accurate methods capable of analyzing and computing encrypted input data and returning the result in encrypted form. In this regard, Zhou, et al. proposed a deep binarized convolutional neural network over encrypted data \cite{New-New-Conf003}. They binarized input data and weights of the model. Additionally, all operations in the model were changed to bit-wise operations. Using this, they achieved a speedup of 6.3 times that of current schemes. Further, in \cite{New-New-Conf004} the authors of been performed cryptanalysis using neural networks and achieved hit rates of 99.5\% with a two-level neural network when classifying file type of stream ciphers in depth using neural networks.
In another work, Badawi, et al. created a homomorphic convolutional neural network that employed GPUs to perform calculations on encrypted data quickly \cite{Crypt-In-AI-Jour034}. They evaluated two CNNs to classify the MNIST and CIFAR-10 datasets homomorphically, and their solution achieved a sufficient security level. Similarly,  Pradeepthi, et al. use ML to identify the encryption algorithm of encrypted text with over 90\% accuracy \cite{Crypt-In-AI-Conf049}. Okada, et al. compared three ML algorithms, Naive Bayes, Support Vector Machine, and C4.5 decision tree, for classifying encrypted traffic \cite{Crypt-In-AI-Conf050}. They observed that the support vector machine achieved the highest accuracy at 97.2\%.  In research by Muliukha, et al., they use ML to classify encrypted-SSL session applications \cite{Crypt-In-AI-Conf052}. Moreover in another work, Khan, et al. found that supervised ML performed better than unsupervised ML when processing encrypted data \cite{Crypt-In-AI-Conf054}.

Another application of AI on encrypted data is image recovery. In response to growing security concerns surrounding cloud computing, Kitayama, et al. show that it is possible to extract histograms-of-oriented-gradients features encrypted then compressed images \cite{Crypt-In-AI-Conf051}. Additionally, in VPN connections they were able to classify user actions. T. Hazra developed an image retrieval for encrypted images. Their solution involves creating a database of features and then having the images be reconstructed by a ML algorithm \cite{Crypt-In-AI-Conf053}.

However, it is important to remember that many applications of encrypted traffic classification are malicious. Msadek, et al. show that it is possible to perform IoT fingerprinting attacks via ML \cite{Crypt-In-AI-Conf055}. They train machine-learning algorithms based on selected features extracted from the encrypted IoT traffic to perform the fingerprint attack. Also, Liu, et al. show that it is possible to identify a user's behavior from features extracted from encrypted surveillance video traffic with 94\% accuracy \cite{Crypt-In-AI-Conf056}.

One particular area of interest is video quality prediction from encrypted traffic. Orsolic, et al. show that it is possible to predict the quality of experience for video streaming from only monitoring encrypted network data with 84\% accuracy \cite{Crypt-In-AI-Conf057}. Seddigh, et al. continue the development of MLTAT, a network classification platform, that is now capable of achieving 90\% accuracy when identifying network traffic at 10Gbps \cite{Crypt-In-AI-Conf058}.

In \cite{Crypt-In-AI-Conf059}, the authors proposed ML-based delay-aware UAV detection and operation mode identification over Encrypted Wi-Fi traffic. The suggested framework extracts feature derived only from packet size and inter-arrival time of encrypted Wi-Fi traffic and can efficiently detect UAVs and identify their operation modes. Also, Al-Hababi, et al. were able to perform a man-in-the-middle attack in a lab setting on various social media platforms using encrypted traffic such as YouTube, WhatsApp and Skype \cite{Crypt-In-AI-Conf061}. They were able to classify traffic vs noise with high accuracy for all three platforms. Jackson, et al. were able to identify different classes of requests to an Amazon Echo via random forest ML from encrypted traffic with a 97\% accuracy \cite{Crypt-In-AI-Conf064}. However, the ML model requires more computation to be trained. Alshammari, et al. used several different algorithms to identify SSH traffic in encrypted traffic \cite{Crypt-In-AI-Conf065}. They found that C4.5 was the best suited of their tested algorithms, achieving a tested detection rate of 83.7\% but potentially as high as 97\%.

Leroux, et al. demonstrate that it is possible to identify what encrypted traffic is in an IPsec or TOR tunnel using ML \cite{Crypt-In-AI-Conf066}.
Seufert, et al. use a random forest model to predict video stalling with high accuracy from encrypted network traffic to allow for quality of experience analysis in real-time \cite{Crypt-In-AI-Conf067}. Their model is capable of predicting within 1s.

By and large, one of the most heavily researched areas of CAAI is encrypted traffic. In this regard, He, et al. show that Long Short-Term Memory (LSTM) recurrent neural networks can outperform traditional autoregressive integrated moving average models in identifying encrypted video traffic \cite{Crypt-In-AI-Conf068}. Similarly, in \cite{AI-In-Cryp-Jour087}, the authors considered UAV detection over Encrypted Wi-Fi traffic. This UAV detection method extracts statistical features derived only from packet size and inter-arrival time of encrypted Wi-Fi traffic and can quickly identify UAV types. The evaluation results show that their proposed method can identify the tested UAVs within 0.15-0.35s with high accuracy of 85.7-95.2\%.
Wassermann, et al. developed ViCrypt, which uses ML to analyze encrypted video traffic to predict performance indicators, including but not limited to stalling and video bitrate \cite{AI-In-Cryp-Jour085}. Suh, et al. use fully homomorphic encryption alongside control synthesis. They demonstrate that their methods cause minimal reductions in precision \cite{New-New-Conf0015}.

Mobile encrypted traffic has also been studied. Aceto, et al. propose DL as a viable method for network classification in the rapidly changing mobile network space \cite{Crypt-In-AI-Conf192}. To address the challenges of identifying BitTorrent traffic in TCP tunnels Cruz, et al. developed a recurrent neural network that uses the statistical behavior of the tunnel to identify BitTorrent traffic \cite{Crypt-In-AI-Conf193}. Lu Vu, et al. created a time series analysis technique to identify important features of the packets \cite{Crypt-In-AI-Conf194}. They then used this technique alongside a ML model to identify various encrypted network traffic with an average F1 score of 98\%.

Alternatively, Shen, et al. developed DeepQoE, which is capable of measuring the quality of experience of encrypted network traffic using a convolutional neural network \cite{Crypt-In-AI-Conf195}. This provided a 22\% increase over other state-of-the-art video quality of experience predictors on encrypted traffic at the time.

In 	\cite{Crypt-In-AI-Conf196}, offered a multi-view DL model for encrypted traffic classification, which uses different neural network structures to process different types of features. Likewise,  the authors of \cite{Crypt-In-AI-Conf197} detect anomalies in Encrypted traffic via deep dictionary learning as D2LAD. The experimental results show that their method achieves high accuracy and low resource usage and outperforms the representative state-of-the-art techniques in real-world settings.

Wang, et al. use three interpolation schemes derived from nearest value, bilinear and bicubic to reduce encrypted network packets down to constant size \cite{Crypt-In-AI-Conf199}.

Soleymanpour, et al. provided a  DL method for encrypted traffic classification on the Web \cite{Crypt-In-AI-Conf200}. The authors adopted a cost-sensitive CNN and considered a cost for each classification according to the distribution of classes. The neural network uses these costs for the training process to improve the overall accuracy. Compared to the Deep Packet method, the proposed model obtained higher classification performance (about 2\% on average) based on the empirical results.
Chen, et al. developed PlaidML-HE, which allows for the automatic creation of homomorphic encryption kernels on many different devices by using domain-specific languages \cite{Crypt-In-AI-Conf201}. Furthermore, they show that PlaidML-HE significantly reduces the runtime of MLaaS using it compared to existing implementations of the time.
Aceto, et al. then showed that the application of big data does not improve the time needed for the training phase of traffic classification DL, despite its ability to accelerate other tasks \cite{Crypt-In-AI-Conf202}.

In another research, the authors have discussed a service-centric segment routing mechanism using reinforcement learning for Encrypted traffic \cite{Crypt-In-AI-Conf203}. They proposed a novel service-centric segment routing mechanism using reinforcement learning in the context of encrypted traffic. Their proposal aims to help ISPs decrease the network problems' influence and meet the strict user requirement related toQuality of Experience.

To tackle the rapid evolution of network traffic diversity, the authors of \cite{Crypt-In-AI-Jour062} proposed a DL-based network encrypted traffic classification and intrusion detection framework. This is a lightweight framework with DL for encrypted traffic classification and intrusion detection, termed deep-full-range. Wang, et al. use three DL schemes to create encrypted data classifiers \cite{Crypt-In-AI-Jour063}. They use these data classifiers in their software-defined network home gateway framework, intended to manage smart home networks better.

Moreover, Hou, et al. show that it is possible to predict a user's activity from their encrypted wireless traffic with up to a 99.17\% accuracy \cite{Crypt-In-AI-Conf191}. Additionally, Shen, et al. show that packet length statistics, when combined with clustering methods based on the similarities of the initial incoming and outgoing packets with certain flags from the same website, can be Incorporated into ML models \cite{New-New-Conf0032}. They refer to this process as Bali. Their experimental results show that ML models using Bali can achieve an accuracy of 79.4\%, an improvement over CUMUL and APPS methods.

To sum up, to protect data against security risks, researchers have explored novel ML and DL methods to process privacy-sensitive data over ciphertext.
Due to the use of ML and DL on encrypted data, CAAI has various applications, including identifying encryption algorithms, image recovery, examining the normal and abnormal behaviour of Wi-Fi traffic, and predicting the quality of video streams by monitoring encrypted network data.
	
\subsection{CFAI}

\begin{figure}[h]
	    \centering
	    \includegraphics[width=0.9\linewidth,keepaspectratio]{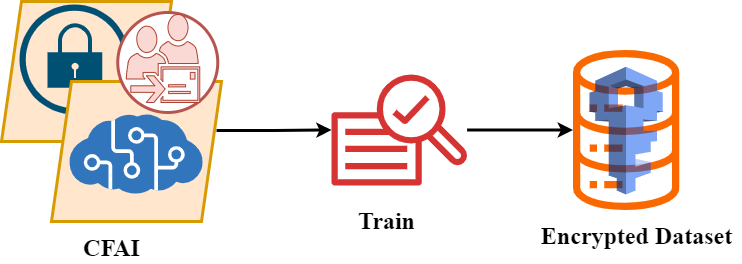}
	    \caption{CFAI Capable of Being Trained on Encrypted Datasets}
	    \label{fig:Crypt-Fri-AI}
	\end{figure}

In the modern era and with the advancement of information and communication technology, AI applications have penetrated many fields.  One of the most critical factors required for the accurate performance of models based on AI is the availability of sufficient data for model training.  However, making data available has legal, security, and privacy issues.
In this regard, CFAI has been of interest to several researchers in recent years.  As we can see in \textcolor{black}{Figure} \ref{fig:Crypt-Fri-AI}, to solve mentioned problems related to data availability, CFAI has made it possible for people to upload their training data in encrypted form and create an encrypted model. For instance, Mokel, et al.  showed that it is possible for a convolutional neural network acting on encrypted data to produce an identical accuracy order to that of a convolutional neural network working on plain data \cite{New-New-Conf006}. They considered training neural networks over encrypted data designing convolutional neural network classifiers when data in their basic form are not accessible because of the privacy policy. Also, the authors of \cite{New-New-Conf007} show that it is possible to achieve a 50x speedup of DL on encrypted data by optimizing the open-source library HElib. Their networks still maintain an acceptable accuracy of 96\% via implementing a Stochastic Gradient Descent (SGD) based neural network training.

In another work, Xu, et al. propose a Crypto Neural Network that uses the emerging functional encryption scheme to allow for training over encrypted data \cite{Train-Jour001}. Their CryptoNN framework employed the emerging functional encryption scheme instead of secure multi-party computation (SMC) or homomorphic encryption (HE) to train a neural network model. Their models' accuracy on the MNIST dataset is similar to other networks. In another work, the same researchers developed another framework that uses encrypted multi-sourced datasets to train neural networks \cite{Train-Jour002}. Their method significantly reduces the time it takes to train a neural network. When tested on the MNIST datasets, their model's accuracy was similar to other neural networks. Also, they found that the depth and complexity of neural networks do not impact the training time despite introducing a privacy-preserving.

F. Gonzalez-Serrano, et al. train ML models on encrypted data by exploiting the partial homomorphic property of certain cryptosystems \cite{Train-Jour003}. They use multiple-key cryptosystems and the impact of quantizing real-valued input data required before encryption. Finally, Bost, et al. developed three classification protocols for
Naive Bayes, decisions trees, and hyperplane decisions 
provide the privacy that relies on a library of building blocks for constructing classifiers securely \cite{Train-Conf001}. They also show the potential for this library to implement other classifiers such as a multiplexer or a face detection classifier.

	
		 	
CFAI, as the third stage of CIAI evolution, in addition to detecting malign encryption and processing encrypted data, is capable of training DL and ML models using encrypted datasets. As can be seen from the results, the accuracy obtained from the models taught on the encrypted dataset is equal to that used in the plain dataset. Moreover, training DL and ML models using encrypted datasets leads to improved data privacy protection.

\subsection{CEAI}

	\begin{figure}[h]
	    \centering
	    \includegraphics[width=0.9\linewidth,keepaspectratio]{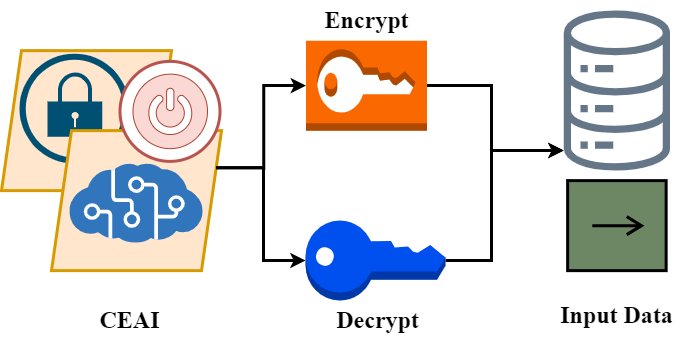}
	    \caption{CEAI Capable of Encrypting and Decrypting Input Data}
	    \label{fig:Crypt-En-AI}
	\end{figure}

Despite the widespread use of AI in cryptography and reducing its vulnerabilities, it still poses dangers. Hackers try to break data encryption by designing complex attacks. They often try to extract plaintext according to the patterns in ciphertext to break the encryption. Therefore, there is a need for a tool that can use complex and unpredictable computations to enhance cryptographic security. In this area, researchers have studied the application of AI in strong cryptography and have achieved excellent results. Various features in AI algorithms such as self-learning, mutual learning, and stochastic learning are helpful encryption. CEAI, which is the next step of the Evolution Stages of CIAI, can reduce the possibility of breaking ciphertexts by producing complex patterns.
 In this method, cryptography can be employed for securing firmware, proprietary IP, and sensitive data, especially in embedded systems. According to \textcolor{black}{Figure} \ref{fig:Crypt-En-AI}, CEAI embeds cryptography into neural networks and other AI models for advanced encryption and decryption on input data. For instance, in \cite{New-New-Jour014}  has proposed cryptography embedded on the convolutional neural network (CNN) accelerator as CREMON model. CREMON is a reconfigurable system with a cryptography reconfigurable processing element (CRPE). 
Moreover, protecting nonvolatile memories from having their trained neural network parameters extracted via a confidentiality attack is becoming increasingly important \cite{AI-In-Cryp-Jour096}. Cai, et al. propose using a runtime encryption scheme and a sparse fast gradient encryption scheme together to ensure confidentiality \cite{New-New-Conf0033}\cite{AI-In-Cryp-Jour096}. They provide variants of the encryption method to better fit quantized models and various mapping strategies. The experiments demonstrate that the NN models can be strictly protected by encrypting a tiny proportion of the weights.

Key exchanging is a major step of cryptography. Extensive studies have been conducted to find a suitable protocol for secure key exchange, such as the Diffie–Hellman key exchange protocol. These traditional key exchange protocols are based on algebraic number theory. Many of these protocols are not suitable for practical use due to security, privacy and efficiency problems. In this way, in \cite{jeong2021neural} authors proposed a novel method based on the Tree Parity Machine(TPM) to share a key between sender and receiver securely. They adopted Vector-Valued Tree Parity Machine (VVTPM), which is more secure and efficient than basic TPM in real-life systems.	Secure key sharing is a significant concern in e-health. To eliminate this concern, the authors in \cite{sarkar2021tree} proposed a novel method based on TPM  to prevent man-in-the-middle attacks in sharing patients' information. This proposed method gained higher performance in terms of protection, reliability, and efficiency than existing key sharing schemes.
	

As observed in all previous evolution stages, AI and cryptography were two separate methods, and AI was used to perform a process on encrypted data. In contrast, this stage is the starting point for AI and cryptography to begin to integrate. In this stage, AI inheritance features, including learning, generating complex patterns, etc., are used to improve cryptography. CEAI allows for more advanced applications of AI models, such as public-key encryption, key distribution, hash generation, etc.

\subsection{CPAI}

\begin{figure}[h]
	    \centering
	    \includegraphics[width=0.9\linewidth,keepaspectratio]{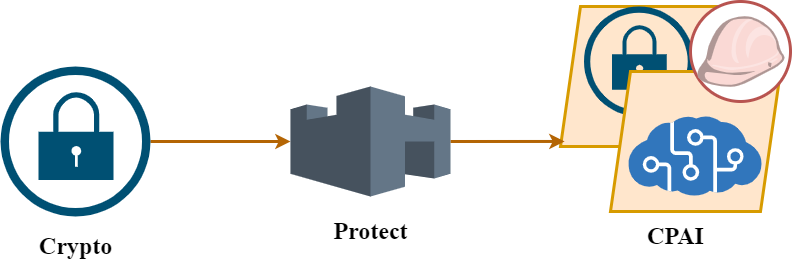}
	    \caption{CPAI}
	    \label{fig:Crypt-Prot-AI}
	\end{figure}

After exploring different approaches to cryptography and AI, the researchers found that the two technologies could not be considered separately to achieve the highest security and performance. This has led to the belief that the intertwining of AI and cryptography can effectively maintain security and privacy. As a result, a new concept called CPAI, which is demonstrated in \textcolor{black}{Figure} \ref{fig:Crypt-Prot-AI} has been formed.

To show how CPAI is formed, we show how crypto undergoes changes to be adoptable by AI, and how AI is upgraded to adopt crypto. 
	
\subsubsection{Crypto moving towards AI}

AI can improve attack detection techniques and create more robust cryptography such as homomorphic encryption. Homomorphic encryption algorithms are a type of encryption algorithm designed to allow mathematical operations to be performed on encrypted data. In this regard, the authors in \cite{Hom-Conf107} developed an expansion on the original TFHE scheme for fully homomorphic encryption. Their new library, WTFHE, is implemented in Ruby to be quick to implement and verify functionality. Their method can be speedily verified
before being implemented as a high-performance library.  Moreover, Arita, et al. created the FHE4GHT scheme for homomorphic encryption, which is capable of computing the greater-than bit of an encryption key \cite{Hom-Conf016}. In other words,  it can homomorphically classify a given encrypted data without decrypting it, using machine-learned parameters.

  In another work, the authors analyzed the robustness of pixel-based image encryption against ciphertext-only attacks \cite{New-New-Conf0013}. They found that the original image could be constructed when the attack has used the images generated using models trained by the same encryption key. However, this was not true for images generated using models trained by multiple encryption keys. Similarly, in \cite{New-New-Conf0014},   the authors proposed a fully homomorphic encryption scheme to calculate the encrypting of the great than bit without needing the secret key.
  

	\subsubsection{AI moving towards Crypto}
		 CPAI uses cryptography to protect AI models by preserving the privacy of the training set and the model's parameters.

		 Homomorphic encryption has proven to be very effective in CPAI. Dalvi, et al. demonstrate that it is possible to protect a vulnerable neural network from a backdoor attack using homomorphic encryption \cite{New-New-Conf0023}. Moreover, Phong, et al. combined DL and additively homomorphic encryption to create a privacy-preserving DL system with tolerable overhead that does not leak information to a server \cite{Hom-Conf092}. Also, in \cite{Hom-Conf226}, the authors improve the preserve privacy in DL via fully homomorphic encryption alongside a convolutional neural network. They argue that this is important in DL models regarding image recognition to protect user privacy. In addition, Pedzisz, et al. created a homomorphic feed-forward network \cite{Crypt-In-AI-Jour066}. With two layers, one exponential and hidden the other for logarithmic prepossessing. This network is intended for adaptive nonlinear filtering. They prove that their proposed network is a competitive alternative to sigmoidal feed-forward networks. Also, in \cite{New-New-Conf0016}, the authors use securing ML engines in IoT applications with attribute-based encryption. Indeed, their proposed approach is passwordless and eliminates the third party in enforcing ML transactions. To estimate their a proposed approach, they perform exploratory ML transactions on the IoT board and estimate the computation time for each transaction. The experimental results indicate that their proposed approach can address security issues in ML computation with low time consumption.
		 
		 In another work, Behera, et al. use homomorphically encrypted input data with a ML model to get encrypted output data from the same model \cite{New-New-Conf0017}. 
		 
		 Chen, et al. suggest a decentralized privacy-preserving DL model for vehicular ad-hoc networks to address privacy and data security concerns in data analysis \cite{New-New-Jour008}. This is done using blockchain, fully homomorphic encryption, and DL. Also, Paul, et al. use homomorphic encryption with a long short-term memory DL model to provide better data privacy when using ML in healthcare setting \cite{New-New-Jour009}. Privacy concerns in robotic systems led Chen, et al. to suggest and demonstrate that a privacy-preserving DL model with homomorphic encryption for use in robotic systems can address these concerns efficiently \cite{New-New-Jour010}.

		The authors of 	\cite{New-New-Conf0019}, focus on DL with chaotic encryption-based secured ethnicity recognition. They propose an ethnicity recognition (ER) model based on convolution neural networks (CNNs). Additionally, a chaotic encryption-based blind digital image watermarking technique is applied to recognized security images. Cover images are used to conceal the recognized image to protect the images from attackers or third parties. Likewise, in \cite{New-New-Conf0020}, it has been considered privacy-preserving distributed learning clustering of healthcare data using cryptography protocols. With the aim of enabling a privacy-preserving version of the clustering algorithm, along with sub-protocols for securing computations, they were able to handle the clustering of vertically partitioned data among the healthcare data providers.
		
		Sirchotedumrong, et al. propose a privacy-preserving scheme for deep neural networks for images encryption that also allows for data augmentation \cite{New-New-Conf0021}. Cantoro, et al. show that various encryption mechanisms increase resilience and fault detection on artificial neural networks \cite{New-New-Conf0022}.

		Another common approach to privacy preservation in the research is using pixel-based encryption. Sirichotedumrong, et al. developed a privacy-preserving scheme for deep neural networks that can be used on images that lack visual information and also have independent encryption keys \cite{New-New-Jour011}. Their experiments show that this scheme defends against ciphertext-only attacks and has high classification performance. Sirichotedumrong, et al. developed a similar privacy-preserving scheme for neural networks that did not require common encryption keys \cite{New-New-Conf0024}.
		
		There are some other potentially promising CPAI in the research. Emmanuel, et al. propose using selected gradients only for encryption in order to reduce overhead costs of privacy-preservation in distributed deep neural networks \cite{New-New-Conf0025}. Sun, et al. present a fully homomorphic encryption scheme that improves on HElib. Their scheme reduces the ciphertext size via relinearization and reduces decryption and modulus noise via modulus switching \cite{New-New-Jour013}. Additionally, we can mention to Li, et al. that they developed a lightweight protection scheme via processing in memory and physically unclonable functions to prevent unauthorized access of data and limit edge devices to protect DL models in the cloud \cite{New-New-Conf0026}. Also, Chi, et al. developed audio encryption techniques to allow audio data to be protected in unprotected environments \cite{New-New-Conf0027}. Additionally, they ensure that the encryption technique is compatible with ML algorithms used in unprotected environments.
		In another work,  Lisin, et al. suggest potential uses of order-preserving encryption for privacy-preserving ML models, such as gradient boosting \cite{New-New-Conf00-15}. Indeed, they considered some existing order-preserving encryption schemes and suggested some cases of ML, where they can be applied to obtain correctly working privacy-preserving ML algorithms. 
		Privacy protection is a major concern in AI. One of the research's solutions to this problem is homomorphic encryption. Li, et al. present HomoPAI, which can securely manage and parallel process data from multiple owners \cite{New-New-Conf0029}. Next, Kuri, et al. present a model for privacy-preserving extreme ML \cite{New-New-Conf0030}. This model can learn from additively homomorphic encrypted data.
		Additionally, Madi, et al. propose the novel Federated Learning Model that is secure from integrity and confidentiality threats from an aggregation server \cite{New-New-Conf0031}. This is done by performing a federated averaging operation using verifiable computing and homomorphic encryption in the encrypted domain. The model's effectiveness is verified on the FEMNIST dataset.

	\section{ CIAI Future Roadmap}\label{Fut}
	
\begin{figure}[h]
	    \centering
	    \includegraphics[width=0.9\linewidth,keepaspectratio]{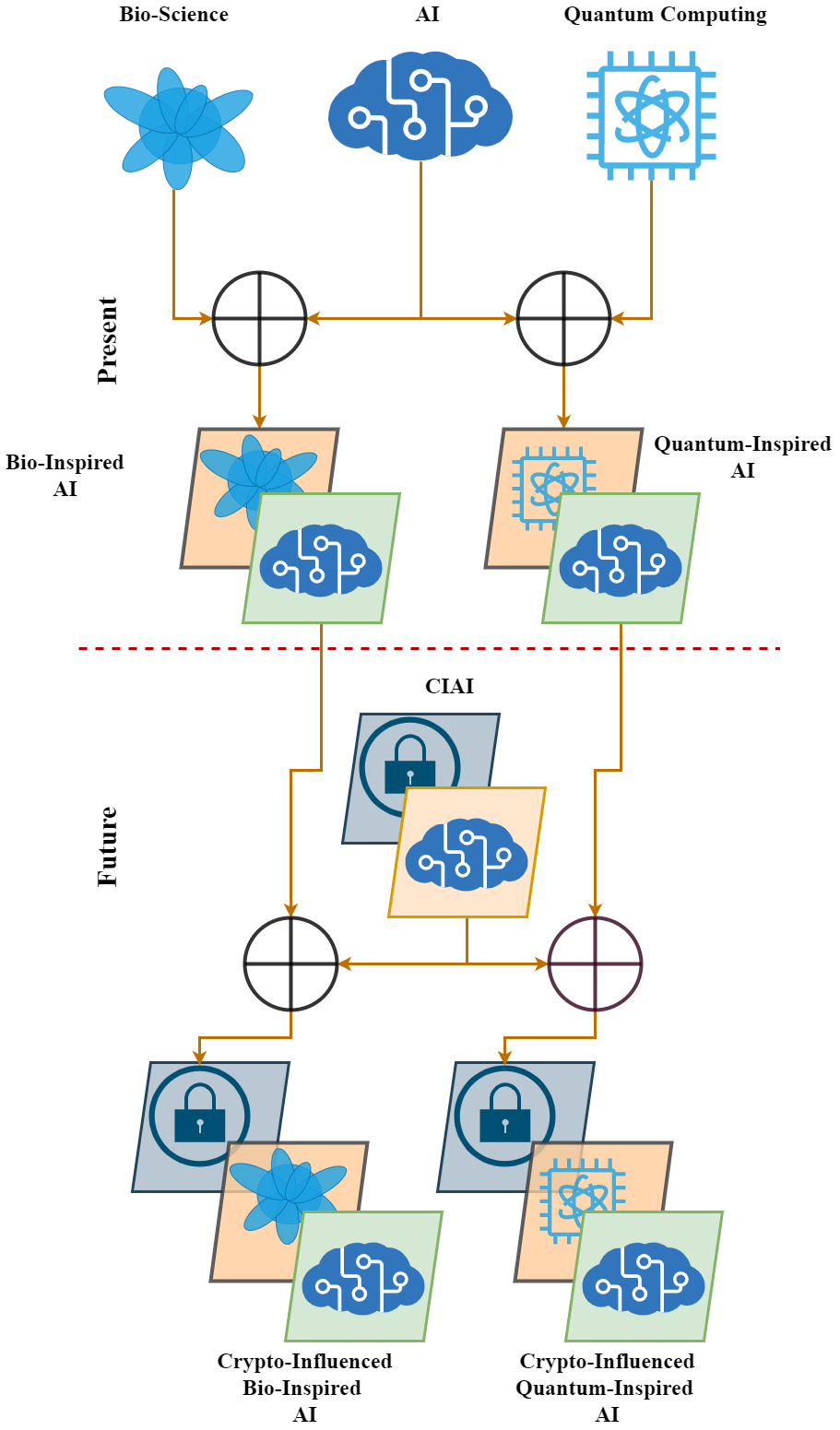}
	    \caption{The Future Roadmap}
	    \label{fig:Future_Movements}
	\end{figure}

	 We anticipate that research on CIAI will move towards crypto-influenced quantum-inspired AI and crypto-influenced bio-inspired AI in the near future. Our reason for such anticipation is the existence of trends towards encrypted AI-based post-quantum cryptography, bio-inspired AI, and quantum-inspired AI, which are discussed in Subsections  \ref{BAI} and \ref{QAI}, respectively.  \textcolor{black}{Figure} \ref{fig:Future_Movements} illustrates the future roadmap of secure quantum AI.  As we can see in the \textcolor{black}{Figure} \ref{fig:Future_Movements}, bio-science and quantum computing are new sciences that have gained researchers' attention due to improving existing challenges in various applications. In this way, researchers seek to combine the feature of these sciences with AI to create a more efficient and more intelligent version of AI. Because of the novelty of bio-inspired AI and quantum-inspired AI, they haven't been used for cryptography by now. In the following, we will explore the potential of bio-inspired AI and quantum-inspired AI for cryptography.
	
\subsection{Bio-Inspired AI}\label{BAI}
 Bio-inspired AI is a new approach to AI which is inspired by evolution. The traditional AI approach has a top-down view and follows the creation approach. In comparison, Bio-inspired AI has a bottom-up view and is inspired by a wide range of biological systems. In this way, Bio-inspired AI is capable of Independent self-organization. Bio-inspired AI has versatile applications, including biorobotics, immune systems, evolutionary computation, genetic algorithm, swarm intelligence, etc \cite{long2020comprehensive}. 
 In the direction of Bio-Inspired AI, we review some related research. Yuan, et al. demonstrate the effectiveness of bio-inspired representation learning for visual attention prediction \cite{Fut-Bio-Jour001}. Their implementation uses bio-inspired ML after feature attraction to combine the high-level semantic features with the low-level contrast features. This allows them to use low-level contrast features to generate a visual attention prediction better than other models. Moreover, Xu, et al. developed a framework for predicting facial aesthetics \cite{Fut-Bio-Jour002}. They first performed experiments to determine regions of interest. Then a convolutional neural network is used to train a set of bio-inspired features.

In another research, Tu, et al. developed a reinforcement learning model to allow for animal movement in robots \cite{Fut-Bio-Jour003}. This model partially or fully takes control of a controller when attempting an animal-like movement such as turning upside down. They then built a hummingbird robot capable of performing a 360 body flip within one wingspan, outperforming other models.
Similarly, the authors of \cite{Fut-Bio-Jour004} use bio-inspired metaheuristic algorithms, such as particle swarm optimization, to optimize many different ML for classifying mail as spam or not. Again, they demonstrate that their results are competitive with other ML models for spam detection.

Lehnert, et al. use retina physiology to create input data for a reinforcement learning network for autonomous navigation \cite{Fut-Bio-Jour005}. They then use auxiliary tasks to generate sparse rewards to allow continuous learning. They found that the auxiliary tasks added robustness to the system in a simulated environment.

The technology for creating bio-inspired AI is improving. Yang, et al. demonstrate that a bio-inspired self-learning method can be used to solve dynamic multiobjective optimization problems in an IoT environment \cite{Fut-Bio-Jour006}. They demonstrate the feasibility of their method on agricultural IoT systems. Their results show that their method outperforms traditional methods, especially for high-dimension problems. Also, Duan, et al. developed BOLE to facilitate the learning of bio-inspired optimization algorithms \cite{Fut-Bio-Jour007}. BOLE is an interactive learning environment based in MatLab that facilitates learning bio-inspired optimization algorithms and is dedicated exclusively to unmanned aerial vehicle path planning with enhancing learning efficiency.

Cipher attacks will get more complex and unpredictable in the future. Therefore, there is a vital need to advance defence methods to improve security. As we discussed, bio-inspired AI creates versatile facilities in different applications. So, bio-inspired AI can be used to enhance cryptography against upcoming cyber threats. For instance, the genetic algorithm can be adopted to design cryptographic protocols, cryptoanalysis and cryptographic primitives \cite{clark2003nature, sadkhan2021proposed}. In another example, some evolutionary algorithms, such as ant colony and swarm, can be used to advance encryption \cite{de2002immune}. Outstanding features of bio-inspired AI  offer a variety of opportunities to cope with emerging cryptography attacks in the future.
	
	\subsection{Quantum-Inspired AI}\label{QAI}
	  Although AI has made much progress in recent years, there are still some challenges, such as reasoning under uncertainty, adaptive ML and neuromorphic cognitive models. Quantum-inspired AI, which means using quantum computing for ML and DL algorithms, is one of the solutions that can help AI level up and overcome these challenges. Quantum-inspired AI can improve AI in terms of improving the decision-making process in complex problems, optimizing ML algorithms, and rapid training of ML models \cite{nawaz2019quantum, dunjko2018machine}.
		 Quantum-inspired AI has begun to show promising results in many applications. As an example in this space, Y. Li et al. use quantum-inspired reinforcement learning to plan the trajectory of unmanned aerial vehicles \cite{Quant-AI-Jour001}. The trajectory planning problem is optimized via the quantum-inspired reinforcement learning (QiRL) approach. Specifically, the QiRL method adopts a novel probabilistic action selection policy and new reinforcement strategy, which are inspired by the collapse phenomenon and amplitude amplification in quantum computation theory, respectively. In another work,
		Wei, et al. applied a quantum-inspired experience relay to traditional deep reinforcement learning  \cite{Quant-AI-Jour002}. Using this, they showed on Atari 2600 games that deep reinforcement learning with a quantum-inspired experience relay had better training efficiency than its competitors.
		
		Masuyama, et al. develop two quantum-inspired multidirectional associative memories \cite{Quant-AI-Jour003}. The first uses a one-shot learning model, and the second uses a self-convergent iterative learning model. They use these models to show the advantages of quantum-inspired multidirectional associative memory, such as stability and reliability.
Moreover, Dong, et al. present a quantum-inspired reinforcement learning algorithm for autonomous robot navigation \cite{Quant-AI-Jour004}. Their simulated experiments show that this approach is more robust than traditional reinforcement learning algorithms.

Finally, Patel, et al. developed a quantum-inspired fuzzy-based neural network for classification between two classes \cite{Quant-AI-Jour005}. Neurons are added to the hidden layer via fuzzy c-means clustering, with the fuzziness parameter being developed from quantum computing. Additionally, this network uses a modified step activation function. Their experiments showed that this network was competitive and superior to current state-of-the-art methods.

According to the quantum-inspired AI developments, we will see its applications in cryptography soon. Each quantum computing and AI have a significant role in improving cryptography. Quantum computing can process large volumes of data at a speed that was not possible with traditional methods. Because AI becomes smarter with more data analyzed, AI algorithms become more accurate, more quickly, with quantum. The popularity of the quantum-inspired AI method for cryptography is not limited to the high speed of learning ML models and data processing. This method can use AI features to more quickly and accurately predict, detect, and neutralize attacks.
	\section{Conclusion}\label{Conc}
	AI has evolved extensively under the impact of cryptography. It started from CSAI, which was primarily used to detect malicious data, to CPAI, which now uses cryptography to protect the privacy of AI models and their datasets. Crypto-AI is now moving into two new research areas, including quantum-inspired and bio-inspired. This paper attempted to determine CIAI evolution stages, including Crypto AI and covered CSAI, CAAI, CFAI, CEAI, and CPAI. Also, a future roadmap was discussed to shed more light on quantum-inspired AI and bio-inspired AI technologies for making a new trend in cryptography. Finally, we discussed how these technologies could turn cryptography into a secure and efficient version.
	

	\bibliographystyle{IEEEtran}
	\bibliography{References}
\end{document}